\documentclass[a4paper,fleqn,usenatbib]{mnras}

\usepackage[T1]{fontenc}
\usepackage{ae,aecompl}

\usepackage{graphicx}	
\usepackage{amssymb}	
\usepackage[usenames,dvipsnames]{color} 

\newcommand{\alm}{a_{\ell m}}

\newcommand{\Cee}{\mathcal{C}}
\newcommand{\Cl}{\Cee_\ell}

\newcommand{\fsky}{f_{\rmn{sky}}}
\newcommand{\Nside}{\textit{Nside}}

\newcommand{\muK}{\,\rmn{\umu K}}
\newcommand*{\code}[1]{\textsc{#1}}

\newcommand{\healpix}{\code{healpix}}
\newcommand{\camb}{\code{Camb}}

\newcommand*{\satellite}[1]{\textit{#1}}
\newcommand{\COBE}{\satellite{COBE}}

\newcommand{\WMAP}{\satellite{WMAP}}
\newcommand{\Planck}{\satellite{Planck}}

\newcommand*{\Planckmap}[1]{\texttt{#1}}
\newcommand{\smica}{\Planckmap{SMICA}}

\newcommand{\common}{\Planckmap{Common}} 

\newcommand{\LCDM}{$\Lambda$CDM}

\renewcommand*{\vec}[1]{\bmath{#1}}
\newcommand*{\unitvec}[1]{\vec{\hat{#1}}}

\newcommand*{\spinfunction}[2]{\,\vphantom{#1}_{#2}#1}

\newcommand*{\ssphYp}[3]{\spinfunction{Y}{+#1}_{#2 #3}}
\newcommand*{\ssphYm}[3]{\spinfunction{Y}{-#1}_{#2 #3}}

\def\lsim{\mathrel{\rlap{\lower4pt\hbox{\hskip1pt$\sim$}}
    \raise1pt\hbox{$<$}}}                
\def\gsim{\mathrel{\rlap{\lower4pt\hbox{\hskip1pt$\sim$}}
    \raise1pt\hbox{$>$}}} 

\newcommand{\sI}{\textit{unconditioned-$\Lambda$CDM}}
\newcommand{\sII}{\textit{\smica-conditioned}}
\newcommand{\sIII}{\textit{Atacama-North}}

\newcommand{\ENorth}{\textit{Ecliptic-North}}
\newcommand{\ESouth}{\textit{Ecliptic-South}}

\newcommand{\cQ}{Q_c}
\newcommand{\cU}{U_c}
\newcommand{\uQ}{Q_u}
\newcommand{\uU}{U_u}

\newcommand{\QU}{\textit{$\langle Q^2 \rangle + \langle U^2\rangle$}}

\newcommand{\iimag}{\rmn{i}}

\title[CMB-S4 and the Hemispherical Variance Anomaly]{CMB-S4 and the Hemispherical Variance Anomaly}

\author[M. O'Dwyer, C.J. Copi, L. Knox  and G.D. Starkman]
{
M\'arcio O'Dwyer$^{1,2}$\thanks{E-mail: marcio.odwyer@case.edu},
Craig J. Copi$^{1}$\thanks{E-mail: cjc5@case.edu},
Lloyd Knox$^{3}$,\thanks{E-mail: lknox@ucdavis.edu}
Glenn D. Starkman$^{1}$\thanks{E-mail: glenn.starkman@case.edu}
\\
$^{1}$CERCA/Department of Physics/ISO, Case Western Reserve University,
 Cleveland, OH 44106-7079, USA\\
$^{2}$The Capes Foundation, Ministry of Education of Brazil, Bras\'ilia DF 70359-970, Brazil\\
$^{3}$Department of Physics, University of California,One Shields Avenue, Davis, CA, USA 95616 \\
}

\date{Accepted XXX. Received YYY; in original form ZZZ}

\pubyear{2016}

\begin{document}
\label{firstpage}
\pagerange{\pageref{firstpage}--\pageref{lastpage}}
\maketitle

\begin{abstract}
Cosmic Microwave Background (CMB) full-sky temperature data show a
hemispherical asymmetry in power nearly aligned with the Ecliptic. In real
space, this anomaly can be quantified by the temperature variance in the
northern and southern Ecliptic hemispheres. In this context, the northern
hemisphere displays an anomalously low variance while the southern hemisphere
appears unremarkable (consistent with expectations from the best-fitting
theory, \LCDM). While this is a well established result in temperature, the
low signal-to-noise ratio in current polarization data prevents a similar
comparison. This will change with a proposed ground-based CMB experiment,
CMB-S4. With that in mind, we generate realizations of polarization maps
constrained by the temperature data and predict the distribution of the
hemispherical variance in polarization considering two different sky
coverage scenarios possible in CMB-S4: full Ecliptic north coverage and
just the portion of the North that can be observed from a ground based
telescope at the high Chilean Atacama plateau. We find that even in the set
of realizations constrained by the temperature data, the low northern
hemisphere variance observed in temperature is not expected in
polarization.  Therefore, an anomalously low variance detection in
polarization would provide strong evidence against the hypothesis that the
temperature anomaly is simply a statistical fluke. We show, within \LCDM,
how variance measurements in both sky coverage scenarios are related. We
find that the variance makes for a good statistic in cases where the sky
coverage is limited, however a full northern coverage is still preferable.
\end{abstract}

\begin{keywords}
cosmic background radiation --  cosmology: observations
-- methods: statistical
\end{keywords}



\section{Introduction}
\label{sec:Intro}

The Cosmic Microwave Background (CMB) temperature fluctuations have been
mapped with increasing precision, starting with the full-sky measurements
of the Cosmic Background Explorer (\COBE), launched in 1989, and culminating with the latest \Planck\ data
release \citep{2015arXiv150205956P}. In most respects, these temperature measurements have
been found to be in excellent agreement with statistical predictions of the
standard cosmological model, inflationary Lambda Cold Dark Matter (\LCDM).
However, some features of the observed CMB temperature fluctuations have
been found to be extremely unlikely within \LCDM. Among these statistical
anomalies \citep[for reviews of anomalies see][]{2010AdAst2010E..92C,2011ApJS..192...17B,2015arXiv150607135P,2015arXiv151007929S} is the Hemispherical Variance Anomaly. Originally
noticed in the first-year \WMAP\ \citep{2003ApJ...583....1B} data, the
feature was originally described as an asymmetry in power 
between hemispheres, nearly best aligned with the northern and southern
Ecliptic hemispheres  \citep{Eriksen:2003db}.

This hemispherical anomaly has been quantified and verified with several
different statistical measures across a variety of angular scales. Some
methods work in real space making use of the N-point correlation function
or the variance whereas other methods work in harmonic space and measure
the anomaly through the angular power spectrum \citep[for a review of
  methods and their application to the latest \Planck\ data see][Sections
  5.2.2 and 6]{2015arXiv150607135P}.
Many of these measures rely on fitting
a dipole modulation to maps of localized estimates of the angular power spectrum.
Another
method, first proposed in \citet{2014ApJ...784L..42A}, consists of fitting
a dipole to a local variance map of the temperature. Comparing with
realizations of the best-fitting \LCDM\ model, it was found that the amplitude
of the dipole in the observed temperature variance map is anomalously large
-- only one in a thousand \Planck\ Full Focal Plane (FFP) simulations
\citep{2015arXiv150906348P} has a dipole amplitude as large as the one found
in the data.

Even though this hemispherical anomaly seems consistent with a dipole
modulation \citep{2015PhRvD..92f3008A}, and the implications of
phenomenological models of dipole modulation on polarization have been
explored \citep{2015JCAP...05..015N}, methods that consider each hemisphere
separately show that the northern Ecliptic hemisphere, in particular,
displays anomalously low power while the southern Ecliptic hemisphere
appears to be consistent with \LCDM\ predictions
\citep{2015arXiv150607135P}. Within the context of the variance, this can
be seen by treating the pixels in each hemisphere separately. It has been
shown \citep{2014ApJ...784L..42A} that only four in one thousand FFP
simulations have lower Ecliptic north variance, while the south shows no
significant deviation from theoretical expectations. Additionally, neither
the difference in variance between the hemispheres nor their ratio appears
to be significantly different from the simulations.  For this reason we
refer to this feature as the Hemispherical Variance Anomaly rather than
the Hemispherical Power Asymmetry, as it is usually called.

While CMB temperature fluctuations have been exhaustively studied, current
polarization maps with large enough sky coverage to precisely characterize
large-angle correlations suffer from low signal-to-noise ratios.  This will
change with CMB-S4, a possible future ground-based `stage 4' CMB survey to follow those being currently deployed (`stage 3'). CMB-S4 is expected to consist of a group of telescopes at the South Pole, in the high Chilean Atacama plateau, and possibly at a northern-hemisphere
site.  High-fidelity polarization measurements will lead to a CMB
polarization map with unprecedented precision covering a significant
portion of the sky.  Especially important for our purposes here, it will be possible to test in
polarization for analogues of some of the large-scale anomalies seen in the
temperature data.

Perhaps the most economical explanation for the temperature anomalies is that they
are all  merely statistical fluctuations -- the so-called `fluke hypothesis'.
The detection of additional anomalies in polarization could be evidence
against this hypothesis.
However, in
order to avoid the concerns about \textit{a posteriori} statistics that
have muddied the waters surrounding the temperature anomalies, it is
crucial to predict in advance the probability distribution functions of a
small number of test statistics for polarization.  Ideally, these should be
chosen on the basis of their ability to discriminate among predictive
physical models for the anomalies, especially in the context of likely
future observations.  However, given the current absence of any such models
for most anomalies, 
one may have to settle for statistics that can be used
to test the fluke hypothesis.  

Even in the absence of specific physical models, one may have reasonable
expectations of properties of such models that could be tested. For
example, it would be reasonable to expect that any physical mechanism that
reduces the variance of the temperature fluctuations in some region of the
sky, would also reduce that of the polarization fluctuations (or at least
the $E$-mode fluctuations which dominate the polarization) in that same
region.  These reasonable expectations should therefore be kept in mind
when designing statistics.

In this work we present a statistic to test for the hemispherical variance
anomaly in upcoming CMB-S4 polarization data. We apply the statistic to
temperature and find anomalously low temperature variance in accordance
with previous results in the literature. Within \LCDM, we generate $E$-mode
polarization realizations constrained by the observed temperature map and
calculate the expected hemispherical variance distributions. We find that
the low Ecliptic north temperature variance is not expected to manifest
itself in the polarization. We analyze the impact of sky coverage on a
possible future measurement by considering the cases of a (terrestrial)
southern hemisphere telescope with partial northern hemisphere coverage
and that of full northern hemisphere coverage.

\section{Hemispherical Variance Method - Temperature and Polarization}
\label{sec:Method}

The variance method \citep{2014ApJ...784L..42A} provides a particularly
simple way of quantifying the difference between the northern and southern
hemisphere. One simply calculates the variance of the pixels in each
hemisphere. It is also well adapted to situations where one is expecting to
have only partial sky coverage, such as is required for the removal of
contaminated pixels (masking) and would certainly be the case for the
northern Ecliptic hemisphere within CMB-S4 in the absence of a (terrestrial)
northern-hemisphere site.

Given that the north-Ecliptic hemisphere displays anomalously low variance
in temperature, one might naturally wonder if the same is to be expected in
polarization.  This question has two parts.  First, given that
polarization and temperature are correlated in \LCDM, does the low variance
in northern temperature imply a low variance in northern polarization?
Second, can CMB-S4 measure the variance in northern
polarization well enough to assess whether that variance is anomalous, and
does the answer to that depend on having a northern-hemisphere site?

A technical issue to address is how best to generalize the hemispherical
variance statistic to polarization given that $Q$ and $U$ are individually
coordinate dependent.  One simple solution is to consider $Q + U$ instead.
Since in \LCDM\ $Q$ and $U$ are uncorrelated, the expected variance of $Q+U$
is a rotationally invariant quantity given by \QU\, where the angled
brackets denote the ensemble average.

\subsection{Constrained realizations of $\bmath{Q}$ and $\bmath{U}$}

From observations by \WMAP\ and \Planck, we know the temperature
fluctuations on the full sky. Thus, given the best-fitting theory, which
provides us the correlations between $T$, $Q$, and $U$, we can generate
full-sky realizations of $Q$ and $U$ constrained by these temperature
measurements. The specifics of how to generate these constrained
realizations are shown in Appendix A of \cite{2013MNRAS.434.3590C}.
Such a set of constrained polarization realizations provides the correct
theoretical posterior probability distribution for comparison to future data.

Following the standard treatment, fluctuation maps are represented using
the usual spherical harmonic coefficients.  For the temperature these are
the $\alm^T$ in
\begin{equation}
T(\unitvec{n}) = \sum_{\ell m} \alm^T Y_{\ell m}(\unitvec{n}),
\end{equation}
where $\unitvec{n}$ is a unit vector representing the direction of
observation on the sky. For the Stokes parameters of polarization we follow
the convention of \citet{2005ApJ...622..759G},
\begin{eqnarray}
	Q(\unitvec{n}) & = & - \sum_{\ell m} \alm^E X_{1,\ell m}(\unitvec{n})
        + \iimag \alm^B X_{2, \ell m}(\unitvec{n}), \\
	U(\unitvec{n}) & = & - \sum_{\ell m} \alm^B X_{1,\ell m}(\unitvec{n})
        - \iimag \alm^E X_{2, \ell m}(\unitvec{n}), 
\end{eqnarray}
where the $\alm^E$ and $\alm^B$ are the standard $E$-mode and $B$-mode
coefficients and $X_{1, \ell m}$ and $X_{2, \ell m}$ are linear
combinations of spin-2 spherical harmonics defined as $X_{1, \ell m} \equiv
(\ssphYp{2}{\ell}{m} + \ssphYm{2}{\ell}{m})/2$ and $X_{2, \ell m} \equiv
(\ssphYp{2}{\ell}{m} - \ssphYm{2}{\ell}{m})/2$.

Since the polarization is dominated by the $E$-mode signal, we will ignore
the $B$-mode contribution.  This simplifies the $Q$ and $U$ expansions
above and has no discernible effect on our conclusions.   
In \LCDM\ the coefficients $\alm^E$ are linear combinations of
a component correlated with the temperature and an uncorrelated
component. Given this linear nature, $Q$ and $U$ can be split into
correlated and uncorrelated parts,
\begin{eqnarray}
	Q & = & \cQ + \uQ, \\
	U & = & \cU + \uU,
\end{eqnarray}
where the subscript $c$ denotes the correlated piece and $u$ the uncorrelated one.

\subsection{Parameters}
\label{sec:parameters}

In this work we consider multipoles up to $\ell = 600$. 
This somewhat arbitrary choice is motivated by the 
observation that  a measurable dipole modulation in the temperature is 
induced above $\ell\simeq600$ by the effect of our peculiar velocity 
with respect to the CMB rest frame \citep{2015JCAP...01..008Q}. 
An improved statistic for low variance  might employ a different maximum value of $\ell$,
and this possibility will be considered in future work. 
The
corresponding \healpix\footnote{See \url{http://healpix.sourceforge.net}} resolution was chosen to be \Nside=256. For the
temperature data we used the \Planck\ \smica\ full-sky map. The Galactic
portion and point sources were removed in all analyses performed using the
\Planck\ \common\ mask.

To properly compare simulations with the \Planck\ \smica\ map, 
we compensated for smoothing with a $5$ arcminute FWHM Gaussian beam.
The degrading procedure was done via \healpix's \texttt{ud\_grade}
function. When degrading the mask, we match the criteria used in 
the \Planck\ Release-2 analyses \citep{2015arXiv150607135P} where all pixels with values less
than $0.9$ are set to zero, and all others to one. All calculations were
done in Ecliptic coordinates. After degrading, the mask was rotated to Ecliptic coordinates
by using \healpix's \texttt{rotate\_alm}. By doing so it is necessary to
calculate the spherical harmonic coefficients from the real-space binary
mask and, after rotation, transform it back to real space. This process
introduces some ringing. Pixels with value approximately under $0.52$ were set to zero
while the remaining were set to one. This threshold was chosen to keep
$\fsky \approx 0.78$ constant.

The theoretical power spectra, $\Cl^{TT}$, $\Cl^{EE}$, and $\Cl^{TE}$, were
generated with the January 2015 version of \camb\footnote{See
  \url{http://camb.info/}}. The parameters were chosen to match the
best-fitting \LCDM\ model from \Planck\ ($TT$ + lowP + lensing)
\citep{2015arXiv150201589P}.

\subsection{Realizations}
\label{sec:realizations}

Our analyses are based on two different sets of realizations that differ in
how the $\alm^E$ were generated. Both sets contain 50 000 realizations and
are defined as follows.
\begin{enumerate}
	\item In the set we call \sI, the $\alm^E$ were randomly drawn from
          Gaussian distributions with variance $\Cl^{EE}$ given by the
          best-fitting \LCDM\ model.
	\item In the set we call \sII, the $\alm^E$ were generated
          constrained by the $\alm^{T}$ extracted from \Planck\ Release-2
          map \smica\ using the procedure described in Appendix A of
          \cite{2013MNRAS.434.3590C}.
\end{enumerate}

Two sky coverage scenarios are considered for each set:
\begin{itemize}
	\item Full-sky with foreground masked out. In this case we generate
          full skies, remove the Galaxy and point sources using the
          \Planck\ Release-2 \common\ mask, and separately perform our
          calculations on the northern and southern Ecliptic
          hemispheres. These configurations will be referred as
          \ENorth\ and \ESouth, respectively.
	\item The portion of the Ecliptic northern hemisphere that can be
          seen from the Chilean Atacama site with foreground masked as in
          the previous configuration. This sky coverage was calculated
          assuming that a ground base telescope at the aforementioned site
          covers Celestial latitudes from $60\degr$ South to $22.5\degr$
          North, corresponding to Ecliptic northern $\fsky \approx 0.40$. We refer to the set of pixels corresponding to this
          coverage as \sIII. 
\end{itemize}

\section{Results}
\label{sec:results}
\begin{figure*}
\centering
  \includegraphics[width=\textwidth]{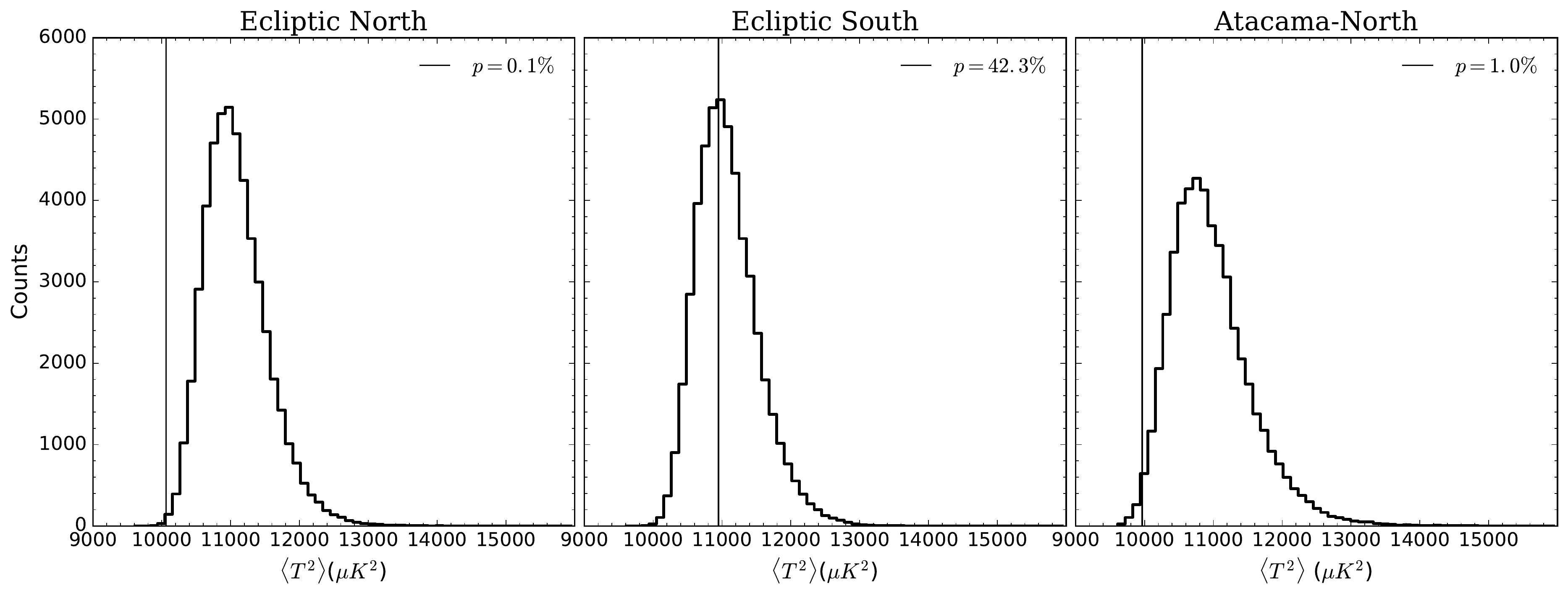}
  \caption{Temperature variance distributions for the \sI\ realizations
    considering three different sky coverage scenarios: pixels in the
    northern Ecliptic hemisphere (\ENorth, left), southern Ecliptic
    hemisphere (\ESouth, middle) and the portion of the northern Ecliptic
    hemisphere that can be seen from the Chilean Atacama site (\sIII,
    right). The vertical lines represent the corresponding values
    calculated from the \Planck\ \smica\ temperature map with $p$-values
    displayed in the plot legends. The \ENorth\ temperature data has an
    anomalously low variance ($p = 0.1$ per cent) when compared with
    \sI\ while the \ESouth\ appears unremarkable ($p = 42.3$ per cent). The
    north variance $p$-value increases to $1.0$ per cent when only the
    \sIII\ pixels are considered. See section \ref{sec:parameters} for
    parameters. }
  \label{fig:Hem_Var_Temperate_N_vs_S_vs_AN}
\end{figure*}

Fig.~\ref{fig:Hem_Var_Temperate_N_vs_S_vs_AN} shows the temperature
variance for \ENorth\ (left), \ESouth\ (middle), and \sII\ (right) for \sI.
Also plotted as vertical lines are the appropriate values calculated from
the \Planck\ \smica\ map. While the \ESouth\ variance has a $p$-value of
approximately $40$ per cent, meaning that $40$ per cent of the theory
realizations had a variance equal to or smaller than the one seen in the
\smica\ map, the \ENorth\ value corresponds to $p=0.1$ per cent. This
result is in agreement with \cite{2014ApJ...784L..42A} and other
hemispherical anomaly studies as summarized in
\cite{2015arXiv150607135P}. When the coverage is limited to \sIII, the
previous full-north $p=0.1$ per cent measurement becomes a $p = 1.0$ per
cent measurement.

\begin{figure*}
\centering
  \includegraphics[width=\textwidth]{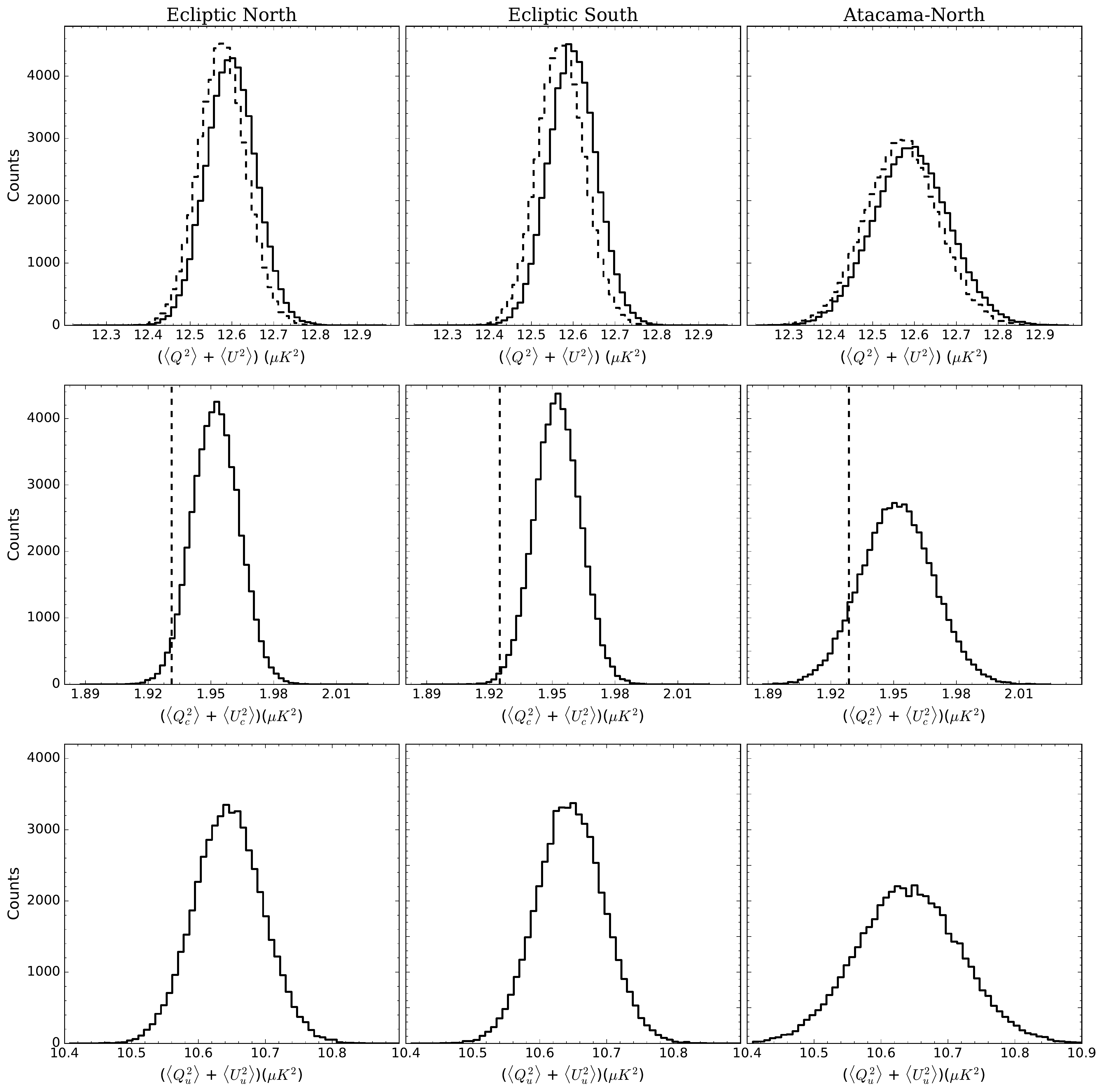}
  \caption{Polarization variance, \QU, calculated for \sI\ (solid) and
    \sII\ realizations (dashed) (See section \ref{sec:realizations} for
    description of realizations). Results displayed in the first column
    correspond to \ENorth\ pixels, while the middle column corresponds to
    \ESouth\ pixels, and the right column to \sIII\ pixels. The second and
    third rows show results for the independent correlated and uncorrelated
    pieces that make up \QU, respectively. As can be seen in the first row,
    the localized lack of variance in the temperature becomes an overall
    slight decrease in variance for polarization (dashed curves).  This
    result is also observed for the correlated piece alone, represented by the vertical dashed line (second
    row). The third column confirms that the smaller sample size of the
    \sIII\ subset of pixels results in broader distributions when compared
    to \ENorth\ and \ESouth.}
  \label{fig:Hem_Var_P_dist_const_vs_unconst_corr_uncorr_with_Atacama_North}
\end{figure*}

In
Fig.~\ref{fig:Hem_Var_P_dist_const_vs_unconst_corr_uncorr_with_Atacama_North}
the polarization results for \QU\ are displayed.  As in
Fig.~\ref{fig:Hem_Var_Temperate_N_vs_S_vs_AN}, the left column shows
\ENorth, the middle one \ESouth, and the right one \sII.  The rows
represent the full polarization variance (top), only the portion of
polarization correlated with the temperature (middle), and the uncorrelated
portion of the polarization (bottom).  In each panel, the solid curve
represents the results from \sI\ while the dashed curve represents those
from \sII. (There is no \sII\ result for the uncorrelated portion of the
polarization.)  As can be seen in the figure, the local lack of power in
the temperature map does not directly translate to a localized lack of
power in \QU. Instead, its effect is only a slight overall decrease in the
variance of the polarization in the \sII\ set.

The \sII\ \QU\ correlated piece
(see the middle row of Fig.~\ref{fig:Hem_Var_P_dist_const_vs_unconst_corr_uncorr_with_Atacama_North}),
does indeed show a suppression of the variance, though less so (in terms of p-value) than the temperature.
Perhaps surprisingly, this is approximately equally true in the northern and southern hemispheres. 
This can be attributed to the $\ell$-weighting applied to $\alm^T$ when
generating the correlated piece of $\alm^E$ \citep[see][Appendix
  A]{2013MNRAS.434.3590C}.
However, the contribution of the correlated piece of \sII\ \QU\ is small compared to the uncorrelated piece
(third row of Fig.~\ref{fig:Hem_Var_P_dist_const_vs_unconst_corr_uncorr_with_Atacama_North}).

\begin{figure}
\centering
  \includegraphics[width=\columnwidth]{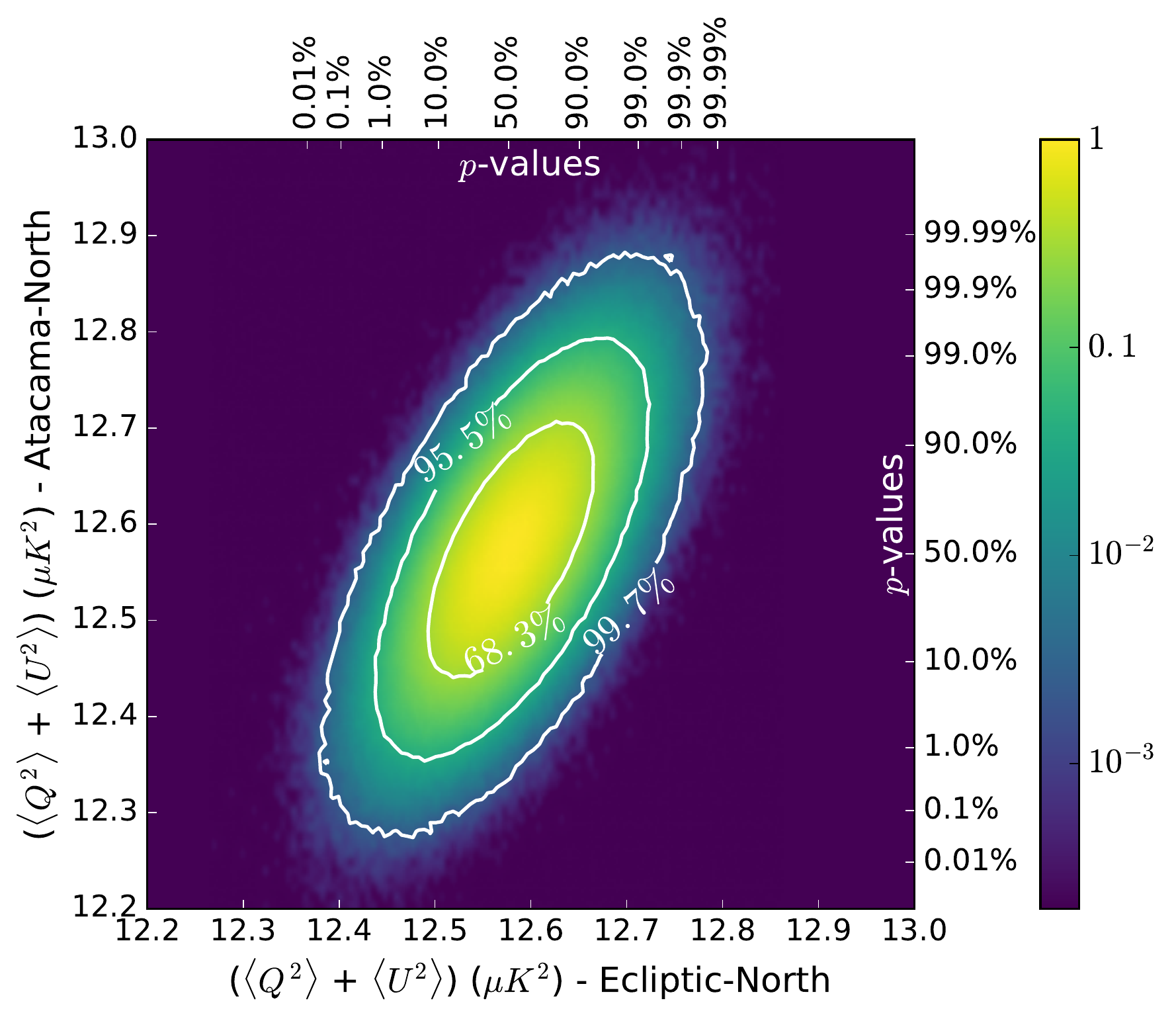}
\caption{Relationship between variance measurements considering the full
  \ENorth\ sky coverage and the partial \sIII\ coverage from an extended
  \sII\ set of realizations (see section \ref{sec:realizations} for
  realization and coverage details). The data are displayed as a density
  plot on a logarithmic scale normalized to unity at the maximum
  height. Confidence curves, defined by constant heights corresponding to
  $68.3$, $95.5$, and $99.7$ per cent of the fraction of points enclosed,
  are overlaid. The $p$-values corresponding to a range of variance
  measurements are shown on the opposing axis.}
  \label{fig:Constrained_P_ecliptic_lmax600_North_vs_Chile_contour}
\end{figure}

The results for \sII\ \QU\ make it clear that under the fluke hypothesis
the polarization is not expected to exhibit the low northern hemisphere
variance that is seen in the temperature data.  
Similarly it would predict no north-south hemispherical asymmetry in the variance.
Thus, any detection in next
generation polarization maps would be considered as evidence against the
fluke hypothesis. Furthermore, if the \ENorth\ polarization variance data
were to suffer a fractional change similar to the one seen in the
temperature, of around $10$ per cent when compared to \ESouth, the
resulting variance of approximately $11.5 \muK^2$ would fall well below the
predicted distribution, even with only partial northern sky coverage. (Also
see Fig.~\ref{fig:Constrained_P_ecliptic_lmax600_North_vs_Chile_contour} .)

The results for the portion of the north Ecliptic sky seen from the Chilean site (\sIII) are displayed in the third column of Fig. \ref{fig:Hem_Var_P_dist_const_vs_unconst_corr_uncorr_with_Atacama_North}. We see that, as expected, for a given measured variance that is smaller than the mean value, the $p$-value in \sIII\ will be larger than the $p$-value in the \ENorth\ as the distributions have the same mean and the former is wider. For any assumed measured variance we could thus work out how much stronger our ability would be to detect anomalous behavior given observations over the whole \ENorth\ rather than just \sIII. However, we only see this if we make the strong assumption that the observed variance in \ENorth\ is the same as the observed variance in the \sIII. We would like to relax this assumption. If we had an actual model that explains the temperature variance anomaly, we could use that model to construct the conditional probability distribution that relates \sIII\ variance to \ENorth\ variance, and thus construct a joint distribution of $p$-values from which one can infer the probability distribution on \ENorth\ given a $p$-value in \sIII\ or the $p$-distribution on \sIII\ given a value in \ENorth. These conditionals would then quantify the benefits of measuring the whole \ENorth\ vs. measuring just \sIII. Of course we do not have such a model. We therefore once again use the only model we do have to obtain some guidance: \LCDM. Specifically, we assume that the fluctuations about some mean variance are the same in our hypothetical anomaly model as in \LCDM. With that assumption, we can calculate the desired joint probability distribution shown in Fig. \ref{fig:Constrained_P_ecliptic_lmax600_North_vs_Chile_contour}. The
data are displayed as a logarithmic density plot normalized to unity at its
peak. Three confidences curves, with probability corresponding to the
fraction of points within the curve defined by a constant height, are
overlaid. We use an extension of the \sII\ set with $4\,500\,000$
realizations. For each axis, a range of $p$-values is placed on the
opposing side of the frame. From this one can deduce how a measurement in
\sIII\ with a certain $p$-value translates into the expected range of
measurements the theory predicts for the \ENorth\ coverage scenario or vice versa. 
For if we expect a $p$-value of $0.1$ per cent on \ENorth, this would lead to 
a measured p-value on \sIII\ up to a few per cent.

We highlight that we have also studied the impact of sky
coverage on the temperature variance utilizing the main method described in
\cite{2014ApJ...784L..42A} where a dipole was fit to a local variance map,
and the statistics of the dipole amplitude were studied. When comparing
methods, simply calculating the variance of the pixels in a hemisphere is
particularly good when dealing with small sky coverage. For the specific
case of the \sIII\ pixels, we constructed the local temperature variance
map on $6\degr$ radius disks. From this we found $p \approx 20$ per cent, a
substantial decrease in significance. Further, the hemispherical variance
method has the added advantage of not depending on an arbitrary choice of
disk size.

\section{Conclusions}

Using the \Planck\ Release-2 \smica\ temperature map and the best-fitting
\Planck\ (TT + lowP + lensing) parameters \citep{2015arXiv150201589P}, we
have predicted the probability distribution for the variance of the
$E$-mode polarization for the full northern Ecliptic sky (\ENorth) and the
portion of it that could be observed from a ground based telescope at the
high Chilean Atacama plateau (\sIII). The latter sky coverage scenario was
motivated by the next generation ground-based CMB experiment, CMB-S4, which
will most likely include a telescope at the aforementioned site.

From our set of $E$-mode realizations constrained by the observed temperature
fluctuations, we have found that  within \LCDM\ the anomalously low \ENorth\ variance
observed in the temperature results in only a slight overall decrease in the 
mean of the polarization variance in both the northern and southern, 
when compared with a set of unconstrained realizations.
This decrease is small compared to the widths of the distributions.
 Neither is any significant asymmetry  predicted between
the Ecliptic hemispheres. 
Therefore a future measurement of polarization variance over both northern and southern 
Ecliptic skies should provide a good test of the fluke hypothesis.

To address the issue of a possible limited northern Ecliptic sky coverage,
we compared how a measurement on the full north relates to the piece
observed from the Chilean site. 
The temperature data show a suppression of around $10$ per cent in the
northern Ecliptic variance when compared to the southern Ecliptic
value, or the theoretical mean.  Lacking a physical model for this suppression,
we cannot predict what the expected signal of such a model would be.
However,  given that the polarization distributions
are much narrower than those for the temperature, a mechanism resulting in
a similar fractional suppression of $10$ percent, 
would fall well below the predicted curve ($p$-value $\ll 0.1$ per cent) for
either Atacama-North or the full North-Ecliptic sky, 
and would be easily detectable even for just \sIII.

For a less extreme suppression of the polarization variance, the
advantages of larger northern sky coverage are clearer.
In the case of CMB temperature  the
observed full northern sky $p$-value of $0.1$  per cent becomes a less extreme $p$-value of $1.0$
per cent on the partial northern sky.  
For polarization, we can predict the probability distribution functions in LCDM
given the observed temperature map.
As expected full-Northern variance and  Atacama-Northern variance are correlated:
their probability distributions have the same mean, and low full-Northern variance predicts low Atacama-Northern variance, and vice versa.
However, also as expected, reduced sky coverage reduces the significance (increases the p-value) of a measurement of suppressed variance,
and we have quantified this.
For example, a p-value of $0.1$ percent on the North-Ecliptic sky 
corresponds to a range of p-values on Atacama-North that can easily reach
up to a few percent.  This suggests that a larger northern sky coverage would increase
the utility of CMB-S4 for probing CMB anomalies, and the hypothesis that they are 
merely a statistical fluke.

\section*{Acknowledgements}
GDS and MO'D are partially supported by Department of Energy grant
DE-SC0009946 to the particle astrophysics theory group at CWRU.  MO'D is
partially supported by the CAPES Foundation of the Ministry of Education of
Brazil.  Some of the results in this paper have been derived using the
\healpix\ \citep{2005ApJ...622..759G} package. This work made use of the High Performance Computing Resource in the Core Facility for Advanced Research Computing at Case Western Reserve University.

\bibliographystyle{mnras}
\bibliography{s4powerasymmetry.bib} 





\bsp	
\label{lastpage}
\end{document}